\documentclass[aip, jcp, amsmath,amssymb, reprint, footinbib]{revtex4-1}

\usepackage{graphicx}
\usepackage{dcolumn}
\usepackage{bm}
\usepackage{textcomp}
\usepackage[mathlines]{lineno}

\usepackage{color} 

\draft 

\newcommand{\bra}{\left\langle}
\newcommand{\ket}{\right\rangle}

\newcommand{\p}{\partial}

\newcommand{\pder}[2]{\frac{\partial #1}{\partial  #2}}

\renewcommand{\d}{\mathrm{d}}

\renewcommand{\vec}[1]{\mbox{\boldmath $#1$}}

\newcommand{\Tm}{k_{\mathrm{B}}T}

\begin{document}


\title{Go-and-Back method: Effective estimation of the hidden motion of proteins from single-molecule time series} 

\author{Makito Miyazaki}
\email{miyazaki@chem.scphys.kyoto-u.ac.jp}
\affiliation{Department of Physics, Graduate School of Science, Kyoto University, Kyoto 606-8502, Japan}
\altaffiliation[Also at ]{Department of Physics, Graduate School of Science, the University of Tokyo}

\author{Takahiro Harada}
\email{harada@phys.s.u-tokyo.ac.jp}
\affiliation{Department of Physics, Graduate School of Science, The University of Tokyo, Tokyo 113-0033, Japan}

\date{\today}

\begin{abstract}
We present an effective method for estimating the motion of proteins from the motion of attached probe particles in single-molecule experiments. The framework naturally incorporates Langevin dynamics to compute the most probable trajectory of the protein. By using a perturbation expansion technique, we achieve computational costs more than three orders of magnitude smaller than the conventional gradient descent method without loss of simplicity in the computation algorithm. We present illustrative applications of the method using simple models of single-molecule experiments and confirm that the proposed method yields reasonable and stable estimates of the hidden motion in a highly efficient manner.
\end{abstract}


\maketitle 

\section{Introduction}
In single-molecule experiments of molecular motors, it has been a widely adopted strategy to visualize continuous stepwise motion by attaching a large probe particle.\cite{Svoboda:1993, Noji:1997, Rief:2000, Yasuda:2001, Greenleaf:2007} Recently, this technique has also been put into use for monitoring conformational changes in proteins that stochastically switch between two or more metastable states.\cite{Shiroguchi:2007, Shiroguchi:2011} Compared to using a fluorescent dye, using a probe particle has the following advantages: First, advances in technology now enable the monitoring of the particle at ultra-high temporal and spatial resolutions of up to $9.1$ \textmu s\cite{Ueno:2010} and $0.1$ nm,\cite{Nugent-Glandorf:2004, Greenleaf:2007} respectively. Second, the particle can be manipulated under optical microscopes, which provides insights into single-molecule mechanics\cite{Rief:2000, Smith:2001, Uemura:2003, Itoh:2004, Watanabe-Nakayama:2008} and energetics.\cite{Liphardt:2002, Toyabe:2010} However, one problem that remains in this method is that the observed motion of the probe particle does not precisely reflect the protein motion. For typical experiments, since the probe is large and loosely connected to the protein, the motion of the probe is usually delayed. To study protein dynamics in detail, the motion and the physical parameters of the protein from the observed trajectory of the probe particle must be estimated.

To date, there has been systematic effort in developing both theoretical and numerical frameworks to determine a discrete state model of proteins from single-molecule fluorescent spectroscopy.\cite{Geva:1998, Berezhkovskii:2000, Cao:2000, Witkoskie:2004a, Flomenbom:2005, Gopich:2006} The framework was recently combined with Bayesian statistics,\cite{Witkoskie:2004b} which allows to analyze the entire sequence of single-molecule data, and it was experimentally demonstrated that the method yields more reliable estimation than the conventional correlation analysis.\cite{Witkoskie:2008} By the use of entire time series, a method to extract effective energy landscape has also been developed recently.\cite{Baba:2007, Baba:2011}

For the time-series analysis of probe particles, there are several numerical approaches to estimating the underlying stepwise trajectories of the molecular motors from the motion of probe particles, and these have been applied to various kinds of experiments.\cite{Chung:1991, Nan:2005, Kerssemakers:2006, Carter:2008, Bozorgui:2010} However, to the best of authors' knowledge, all of these approaches attempt to discretize the observed trajectories, i.e., the probe trajectories into several discrete states. More importantly, these approaches do not incorporate the dynamics of the entire system, namely, thermal fluctuations of both the protein and the probe, and the response delay of the probe motion. Instead, a sampling technique of the reaction pathway in continuous space, so-called transition path sampling (TPS),\cite{Dellago:1998, Bolhuis:2002} is based on Langevin dynamics. However, this requires significant computational cost to search for the dominant pathways, making it inefficient in the presence of multiple reaction pathways.\cite{Autieri:2009} Although an effective method of estimating dominant reaction pathways (DRPs) has recently been developed,\cite{Faccioli:2006, Sega:2007, Autieri:2009} it is not straightforwardly applicable to the analysis of time series data because the method performs the path sampling using not constant time steps but constant displacement steps.

In the present article, we consider a Langevin system that consists of two Brownian particles (one is visible and the other is hidden) connected with each other. On the basis of this model, we propose a method to efficiently estimate DRPs of the hidden variable from the trajectories of the visible variable. Although the model is very simple, it can be considered as a crude description of single-molecule experimental setups under appropriate approximations. We assume that the model and all the parameters have been determined and focus on the estimation of the DRP of the hidden variable. As will be shown later, even though the model is already given, finding the most probable trajectory of the hidden variable using the trajectory of the visible variable remains non-trivial. For parameter estimation in the presence of hidden degrees of freedom, we have proposed a general framework and the practical utility was demonstrated by a simple Langevin model.\cite{Miyazaki:2011} In this framework, the DRP plays a central role in parameter estimation. Our final remarks in the present article are a discussion of practical application of the proposed method to parameter estimation.

Once the model is given, the path-probability can be expressed in terms of the Onsager--Machlup path probability.\cite{Onsager:1953, Machlup:1953, Hunt:1981} Thus, in principle, we can apply a standard maximum likelihood estimation to the hidden trajectory. However, we find in general that the conventional optimization algorithm requires too high a computational cost to find the DRP. Here, we develop an effective approximation technique with the aid of perturbation theory. The schematic procedure of our method is as follows. First, on the basis of a rough estimate of the DRP, we solve one differential equation in the forward-time direction. Next, by substituting this solution, we solve another differential equation in the reverse-time direction and obtain a better estimate of the DRP. By repeating this procedure, we can systematically increase the accuracy of the estimate. Since we solve these differential equations by alternating between the forward and reverse (backward) directions, we name this algorithm the Go-and-Back method.

In Sec.~II, we introduce a working model and discuss the validity of the model. Then, we explain the problem with the gradient descent method and derive the Go-and-Back method. In Sec.~III, we examine two models of typical single-molecule experiments to investigate the effectiveness of our method.

\begin{figure}[tbp]
\centerline{\includegraphics{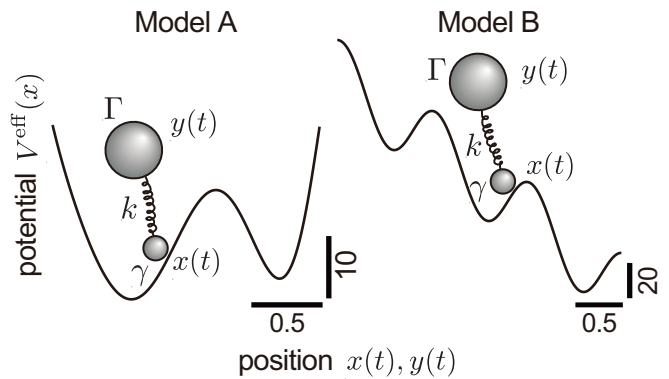}}
\caption{Models for the numerical experiments. Model A: $x(t)$ is stochastically switching between two local minima of $V^{\rm eff}(x)$ by means of thermal noise (equilibrium state). Model B: $x(t)$ is stochastically stepping down the tilted effective potential $V^{\rm eff}(x) \equiv V(x) - fx$, where $V(x)$ is a periodic function and $f$ is the driving force (non-equilibrium steady-state). The parameters in these models are chosen as follows. Model A: $V^{\rm eff}(x)$ is a polynomial function defined as $V^{\rm eff}(x) \equiv \sum_{i=1}^8 a_i x^i$, where $a_1 = 2.57$, $a_2 = 114.4$, $a_3 = -14.0$, $a_4 = -191.9$, $a_5 = -81.3$, $a_6 = 196.4$, $a_7 = 29.1$, and $a_8 = -51.5$.\footnote{In typical experiments, the bimodal distribution of the probe particle is best fitted by two Gaussian functions. Thus, we roughly fitted two Gaussian functions of different means and variances with an eighth-order polynomial function in order to reproduce experimental results.}
Model B: $V(x) \equiv A \sin(2\pi x/l)$, where $A = 20$, $l = 1$, and $f = 40$. In both models, $\gamma = 2$, $\Gamma = 20$, $k = 160$, $\Tm = 4.11$, and $\Delta t = 0.01$.}
\label{f.model}
\end{figure}

\section{Framework}

\subsection{Model}
The molecular structure of protein typically consists of a few hundred of amino acids. Therefore, in general, we need to consider such a huge number of degrees of freedom to study molecular dynamics of proteins. However, recent experimental\cite{Adachi:2007, Toyabe:2010, Hayashi:2010, Shiroguchi:2007, Shiroguchi:2011} studies on motor proteins clarified that only a few degrees of freedom dominate the large-scale conformational changes. In particular case of the rotational molecular motor F$_1$-ATPase, it has been experimentally shown that the energy conservation of the entire system is explained by considering only one-dimensional (rotational) motion.\cite{Toyabe:2010, Hayashi:2010} In addition, normal mode analysis on motor proteins\cite{Cui:2004, Cui:2006, Togashi:2007, Togashi:2010} implies that the low-frequency modes (a few degrees of freedom) correspond to such large-scale motion (\textmu s--ms), and the low-frequency modes are distinct from higher-frequency modes (a huge number of degrees of freedom) that may correspond to the local conformational fluctuations and the catalytic reactions at the active site (ps--ns). Therefore, within the typical time resolution of an optical microscope (\textmu s--ms), the large number of high-frequency modes are eliminated from the dynamics of proteins and thus the dynamics can be approximated by low-dimensional overdamped Langevin equations.

Taking into account the above facts, we consider a Langevin system that consists of two Brownian particles interacting with each other:
\begin{eqnarray}
	\gamma \dot{x} &=& -\p_x[V^{\rm eff}(x) + U(x,y)] + \xi, \label{e.model_x} \\
	\Gamma \dot{y} &=& F(y) -\p_y U(x,y) + \eta, \label{e.model_y}
\end{eqnarray}
where $\gamma$ and $\Gamma$ are the friction coefficients, $\xi(t)$ and $\eta(t)$ are zero-mean white Gaussian noise with variances $2\gamma \Tm$ and $2\Gamma \Tm$, respectively. If $x(t)$ and $y(t)$ are regarded as the dominant degrees of freedom of the protein and the probe particle, respectively, Eqs.~(\ref{e.model_x}) and~(\ref{e.model_y}) can be considered as a crude model of single-molecule experiments.\cite{Julicher:1997, Reimann:2002, Miyazaki:2011} Note that we consider the simplest case where both $x(t)$ and $y(t)$ are one-dimensional variables because typical single-molecule experiments monitor only one-dimensional motion, but the following calculation is straightforwardly extended to higher dimensional $x(t)$ and $y(t)$.

In the present model, $\gamma$ corresponds to the sum of the internal friction coefficient of the protein and the viscous friction coefficient between the protein and the medium. $\Gamma$ corresponds to the viscous friction coefficient of the probe particle. For simplicity, we assume that $\gamma$ and $\Gamma$ are position-independent. $U(x, y)$ corresponds to the energy potential of the elastic linker between the protein and the probe particle. $V^{\rm eff}(x) \equiv V(x) - fx$ is the effective energy potential of the protein, where $V(x)$ corresponds to the energy landscape profile of the protein along the reaction coordinate and $f$ corresponds to the ``driving force'' provided by the catalytic reaction such as ATP hydrolysis. In actual experiments, trapping force or space-constant load is sometimes applied to the probe particle. We also incorporate such an external force into the model equations denoted by $F(y)$. We assume that $U(x,y)$, $V^{\rm eff}(x)$, and $F(y)$ are independent of $t$. 

Here, let us discuss the validity of the working model by using two examples displayed in Fig.~\ref{f.model}. First, we consider that a protein has two chemical states: A and B states, and the protein stochastically goes back and forth between the two states. We suppose that the two chemical states exhibit different conformations and we can observe the switching motion by attaching a probe particle. Here, if the off rates of A(substrate) and B(product) from the protein are slower than the switching rates between the two states and the observation period, the entire system is well approximated as an equilibrium state. In addition, if the switching rates are slower than the global relaxation timescale of the protein structure, namely, the transition rates are well explained by the Kramers' model,\cite{Kramers:1940} the entire dynamics can be modeled by a double-well potential of $V(x)$ with $f = 0$ and $F = 0$ (Fig.~\ref{f.model}a). 

Next, we suppose that a molecular motor translocates in one direction with regular steps by catalytic reactions. For example, a rotary molecular motor F$_1$-ATPase rotates counterclockwise with regular 120$^\circ$ steps by hydrolyzing ATP.\cite{Noji:1997} In typical observation period(several minutes), the entire system is regarded as a nonequilibrium steady-state because the concentration of ATP and the products(ADP and P$_i$) are almost constant in this period. If ATP molecules are abundant in the medium and the rate limiting reaction of each step is the global conformational change of the protein by its thermal fluctuation, the phenomenological model can be described by using a tilted periodic potential (Fig.~\ref{f.model}b).

In this manner, the dynamics of proteins and attached probe particles can be approximated as the simple Langevin model under appropriate conditions. Of course, the present model has several limitations to apply actual experiments due to the simple approximations especially the position-independent $\gamma$ and the time-independent $V^{\rm eff}(x)$. The details will be discussed in Sec.~IV.

In what follows, we assume that $y(t)$ is observed with a sufficiently high temporal and spacial resolution, while $x(t)$ is hidden. We also assume that the entire set of system parameters is given. Then, our final task to consider here is the estimation of the most probable trajectory of $x(t)$ from the trajectory of $y(t)$.

\begin{widetext}

\subsection{Path probability}
We denote the set of the trajectories from time $t = 0$ to $t = \tau$ as $[x, y]$ and the entire set of system parameters as $\vec \Pi = (\Pi_1, \Pi_2, \cdots, \Pi_p)$. Given a value of $\vec \Pi$, we can calculate the path probability $P([x,y] | \vec \Pi)$ as follows.

First, $P([x,y] | \vec \Pi)$ is decomposed into the initial and transition probabilities as
 \begin{eqnarray}
 	P([x,y]|\vec \Pi) = P_{\rm init}(x_0, y_0 | \vec \Pi)P_{\rm tr}((x_0, y_0) \to [x,y] | \vec \Pi),
	\label{e.path}
\end{eqnarray}
where $x_0$ and $y_0$ denote $x(0)$ and $y(0)$, respectively.

Next, the transition probability can be expressed in terms of the Onsager--Machlup path-probability:\cite{Onsager:1953, Machlup:1953, Hunt:1981}
\begin{eqnarray}
	P_{\rm tr}((x_0, y_0) \to [x,y] | \vec \Pi) &=& C \exp[-\beta S([x,y]; \vec \Pi)],
\label{e.transition}
\end{eqnarray}
where $\beta^{-1}\equiv \Tm$ and the action functional $S([x,y]; \vec \Pi)$ is defined as
\begin{eqnarray}
	S([x,y]; \vec \Pi) &\equiv& \frac{1}{4\gamma} \int_0^{\tau} [\gamma \dot{x} + V^{\rm eff}_x(x) + U_x(x, y)]^2 \ \d t - \frac{\Tm}{2\gamma} \int_0^{\tau} [V^{\rm eff}_{xx}(x) + U_{xx}(x, y)] \ \d t \nonumber \\
	&& + \frac{1}{4\Gamma} \int_0^{\tau} [\Gamma \dot{y} - F(y) + U_y(x, y)]^2 \ \d t - \frac{\Tm}{2\Gamma} \int_0^{\tau} [ -F_y(y) + U_{yy}(x, y)] \ \d t,
\label{e.action}
\end{eqnarray}
where the total and partial differentiations are denoted using the same notation such as ${V^{\rm eff}}' \equiv V^{\rm eff}_x$ and $\p_{xx} U \equiv U_{xx}$.
When the trajectory $[x, y]$ is time-discretized by $\Delta t$, the normalization constant becomes $C = [\sqrt{\gamma\Gamma}/ (4\pi \Tm \Delta t) ]^N$, where $\tau \equiv N\Delta t$.

Therefore, if we adopt an appropriate approximation for the initial distribution $P_{\rm init}(x_0, y_0 | \vec \Pi)$,\footnote{If the system is locally equilibrated, we can adopt the Boltzmann distribution.} we can compute the path probability by Eqs.~(\ref{e.path})--(\ref{e.action}).

Once we obtain a concrete expression for the path probability, we can estimate the DRP by a standard maximum likelihood estimation with respect to $[x]$.  In Bayesian statistics, the maximum likelihood estimator (MLE) is included in the maximum {\it a posteriori} (MAP) estimator as a special case. It has been clarified in our previous work that the MAP estimator does not coincide with the true trajectory of $x(t)$.\cite{Miyazaki:2011} However, when the motion of $x(t)$ is stepwise like that of molecular motors, we find that the MAP estimator seems to be a good estimator of the stepping motion of $x(t)$. To maintain consistency with our previous work and also for convenience when considering the application of the Go-and-Back method to parameter estimation, we refer to the MLE as the MAP estimator in the present article, which we denote by $[\hat{x}]$ in the following sections.

\subsection{Gradient descent}
The gradient descent method is the most widely used optimization algorithm. Before proceeding to the Go-and-Back method, let us consider why this standard method is inefficient for the present problem.

To maximize $P([x,y]|\vec \Pi)$ with respect to $[x]$, an initial condition for $[x]$ and a boundary condition are required. Here, we adopt the Dirichlet boundary condition. (For the initial condition, for instance we can adopt $[x] = [y]$.) In this case, $x_0$ is fixed, and thus we avoid having to consider the initial distribution $P_{\rm init}(x_0, y_0 | \vec \Pi)$. Therefore, the maximization of $P([x,y]|\vec \Pi)$ is replaced by the minimization of action functional $S([x,y];\vec \Pi)$ with respect to $[x]$. By introducing a ``virtual time'' $s$ and replacing $x(t)$ with $x(t,s)$, we obtain the MAP estimator $[\hat{x}]$ as the solution of the following partial differential equation in the limit of $s \to \infty$:
\begin{eqnarray}
	\pder{x(t,s)}{s} &=& -\frac{\delta S([x,y];\vec \Pi)}{\delta x} \nonumber \\
	&=& - \left. \pder{W(x,y)}{x}  \right|_{x = x(t,s)} + \frac{\gamma}{2} \frac{\p^2 x(t, s)}{\p^2 t},
	\label{e.relaxation}
\end{eqnarray}
where
\begin{eqnarray}
	W(x, y) &\equiv& \frac{1}{4\gamma}[V^{\rm eff}_x(x) + U_x(x,y)]^2 - \frac{\Tm}{2\gamma}[V^{\rm eff}_{xx}(x) + U_{xx}(x,y)]\nonumber \\
	&&+ \frac{1}{4\Gamma}[ - F(y) + U_y(x,y)]^2 - \frac{\Tm}{2\Gamma}[ -F_y(y) + U_{yy}(x,y)].
	\label{e.w}
\end{eqnarray}
Here, if we consider $s$ in Eq.~(\ref{e.relaxation}) as a ``real'' time $t$, and $t$ in Eq.~(\ref{e.relaxation}) as a spacial coordinate $x$, the form of Eq.~(\ref{e.relaxation}) is similar to the time-dependent Ginzburg--Landau (TDGL) equation for non-conserved systems.\cite{Gunton:1983} When $W(x, y)$ takes the form of a multi-well function, it is known in general that, after the fast relaxation of the local fluctuation, the global relaxation (the kink motion) takes a very long time.\cite{Kawasaki:1982, Carr:1989} Therefore, in the presence of the hopping motion of $x(t)$ between multiple local minima, the gradient descent method may require a large computational cost.

\end{widetext}

\subsection{Go-and-Back method}
As mentioned in the previous section, the gradient descent method requires a large computational cost. On the basis of perturbation theory and making the proper assumption that $\gamma/\Gamma \ll 1$, we solve this problem as shown below.

To solve the Euler--Lagrange equation
\begin{eqnarray}
	\frac{\delta S([x,y];\vec \Pi)}{\delta x} = \pder{W(x,y)}{x} - \frac{\gamma}{2} \ddot{x} = 0,
	\label{e.E-L}
\end{eqnarray}
we decompose the equation into two first-order ordinary differential equations by introducing $v(t)$ which satisfies 
\begin{eqnarray}
	\gamma \dot x = - [V^{\rm eff}_x(x) + U_x(x,y)] + v.
	\label{e.led}
\end{eqnarray}
 $v(t)$ is a deterministic variable which mimics the random force $\xi(t)$ [See Eq.~(\ref{e.model_x})]. Substituting Eq.~(\ref{e.led}) into Eq.~(\ref{e.E-L}) and using Eq.~(\ref{e.model_y}), we finally get
\begin{eqnarray}
	\gamma \dot{v} &=& [V^{\rm eff}_{xx}(x) + U_{xx}(x,y)] v  - \Tm[V^{\rm eff}_{xxx} (x) + U_{xxx} (x,y)] \nonumber \\
	&& + \frac{\gamma}{\Gamma} U_{xy}(x,y)[U_y(x,y) - U_y(x^*, y)] \nonumber \\
	&& + \frac{\gamma}{\Gamma} U_{xy}(x, y)\cdot \eta^*,
	\label{e.dot_v}
\end{eqnarray}
where $\cdot$ represents It\^{o}-type stochastic calculi\cite{Mortensen:1969} and $x^*$ represents the true position of $x$ at time $t$. Since $[y]$ is realized under the true $[x]$, we must use $x^*$ instead of $x$ in Eq.~(\ref{e.model_y}). 

Here, unknown variables $x^*$ and $\eta^*$ are involved in the third and the fourth terms in the right-hand side of Eq.~(\ref{e.dot_v}). Fortunately, the contribution of these two terms are negligible. First, $[U_y(x,y) - U_y(x^*, y)]$ is statistically close to $0$ if $[x] \simeq [\hat{x}]$. Similarly, $\bra U_{xy}(x, y)\cdot \eta^* \ket = 0$. This calculation is independent of $[x]$. Moreover, considering the typical case of single-molecule experiments, we can naturally assume $\gamma/\Gamma \ll 1$.

Therefore, in principle, we may obtain a good approximate solution of $[\hat x]$ by solving the following differential equations simultaneously:\begin{eqnarray}
	\gamma \dot x &=& -G_x(x,y) + v, \label{e.led_g} \\
	\gamma \dot{v} &=& G_{xx}(x,y)  v - \Tm G_{xxx}(x,y), \label{e.v_g}
\end{eqnarray}
where $G(x,y) \equiv V^{\rm eff}(x) + U(x,y)$. Here, Eq.~(\ref{e.led}) is rewritten as Eq.~(\ref{e.led_g}), and Eq.~(\ref{e.dot_v}) is approximated as Eq.~(\ref{e.v_g}). 

However, a difficulty remains: $G_{xx}(x,y)$ on the right-hand side of Eq.~(\ref{e.v_g}) is positive in the typical case when $x(t)$ is fluctuating around the bottom of the energy potential. Thus, as soon as we numerically integrate Eqs.~(\ref{e.led_g}) and~(\ref{e.v_g}), $v(t)$ will destabilize and finally diverge. In contrast, if we try to solve these equations in the reverse-time direction, $x(t)$ will be instantly destabilized due to the same problem.\footnote{When $x$ is close to the bottom of the effective potential, $-G_x(x, y) \simeq - G_{xx}(x_{\rm b}, y) (x - x_{\rm b})$, where $x_{\rm b}$ represents the bottom position.  Since $G_{xx}(x_{\rm b}, y)$ is positive, Eq.~(\ref{e.led_g}) will be destabilized if the equation is integrated in the reverse-time direction.}

To overcome such an initial value problem, we use that the second term in the right-hand side of Eq.~(\ref{e.v_g}) is small. 
In actual experiments, the data are discretized with $\Delta t$. This time interval is sub-millisecond for typical cases, which is much longer than the timescale of the local relaxation of proteins (ps--ns). Within this coarse timescale, the rough energy landscape is effectively smoothed. In addition, for most times $x(t)$ spends in the potential well. In this period, the second derivative of the effective potential $G_{xx}(x, y)$ becomes dominant and the higher-order derivatives will be negligible. Hence the second term in Eq.~(\ref{e.v_g}) will be small for most times.

We introduce a perturbation parameter $\varepsilon$,\footnote{$\varepsilon$ is a nondimensional parameter, which guarantees the smallness of the second term in the right-hand side of Eq.~(\ref{e.v_g}). After the calculation, we substitute $\varepsilon = 1$ into the result.} and rewrite Eq.~(\ref{e.v_g}) as
\begin{eqnarray}
	\gamma \dot{v} &=& G_{xx}(x,y) v - \varepsilon \ \Tm G_{xxx}(x,y).
	\label{e.v_ge}
\end{eqnarray}
We expand $x(t)$ and $v(t)$ in terms of power series of $\varepsilon$, and we define the $i$-th order approximation as
\begin{eqnarray}
	x_{(i)}(t) &\equiv& \sum_{j=0}^i \varepsilon^j x_{(j)}(t),  \label{e.x_purt} \\
	v_{(i)}(t) &\equiv& \sum_{j=0}^i \varepsilon^j v_{(j)}(t).  \label{e.v_purt} 	
\end{eqnarray}
Note that the present definition of the $i$-th term differs from the usual definition: the $i$-th order approximation includes all terms from orders $0$ to $i$. This definition is crucial to dramatically simplify the algorithm. In addition, we introduce a proper approximation for the zeroth-order term of $v(t)$:
\begin{eqnarray}
	v_{(0)}(t) = 0.
 	\label{e.v_0}
\end{eqnarray}
(For the reason, see Appendix A.) Then, once we adopt appropriate boundary conditions for $x_{(i)}(0)$ and $v_{(i)}(\tau)$, the higher-order approximate solutions of $[\hat{x}]$ can be successively obtained as follows. (For details of the derivation, see Appendix A.)
\newpage
 \textbf{Go-and-Back method:}
 \begin{itemize}
 \item[A.]
 Using $v_{(i)}(t)$, we solve
\begin{eqnarray}
	\gamma \dot{x}_{(i)} &=& -G_x (x_{(i)},y) + v_{(i)} + O(\varepsilon^{i+1})
	\label{e.gb_x}
\end{eqnarray}
\underline{from $t = 0$ to $t = \tau$} and obtain $x_{(i)}(t)$.
 \item[B.]
 Using $x_{(i)}(t)$, we solve
 \begin{eqnarray}
	\gamma \dot{v}_{(i+1)} &=& G_{xx}(x_{(i)},y) v_{(i+1)} - \varepsilon \ \Tm G_{xxx}(x_{(i)},y)\nonumber \\
 	&& + O(\varepsilon^{i+2}).
	\label{e.gb_v}
\end{eqnarray}
\underline{from $t = \tau$ to $t = 0$} and obtain $v_{(i+1)}(t)$.
\item[C.]
Alternate between Step A and Step B. 
\end{itemize}


\begin{figure*}[t]
\centerline{\includegraphics{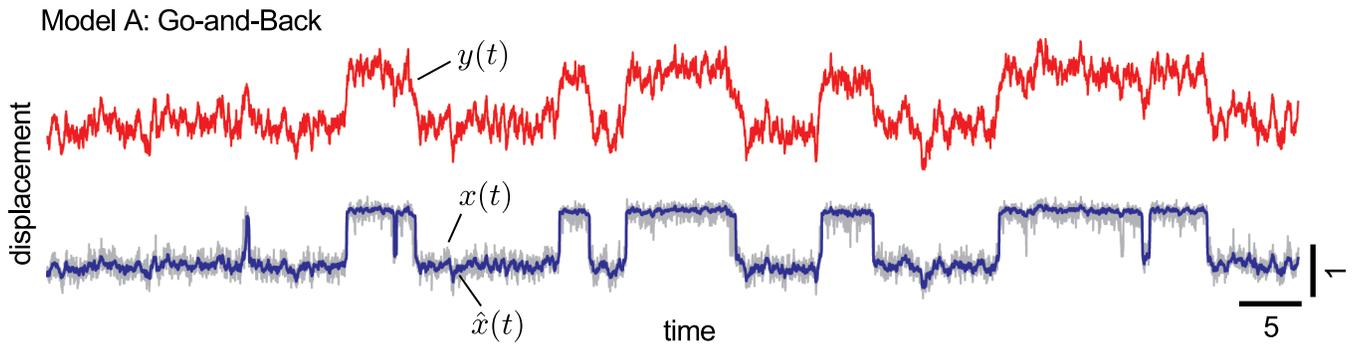}}
\caption{Estimation result of model A by the Go-and-Back method. (top) The trajectory of $y(t)$ [red line]. (bottom) The true trajectory of $x(t)$ [gray line], and the estimated trajectory of $x(t)$, denoted as $\hat{x}(t)$ [blue line]. The iteration number of the optimization process is $i=50$, and the data length is $\tau = 100$. For the present parameter setting, the relaxation time scale of $x(t)$ is shorter than $\Delta t$. Therefore, we use linear-interpolated $[y]$ with the step size $\Delta t/10$ to stably integrate Eqs.~(\ref{e.gb_x}) and (\ref{e.gb_v}).}
\label{f.tc_dw_gb}
\end{figure*}

\begin{figure*}[tbp]
\centerline{\includegraphics{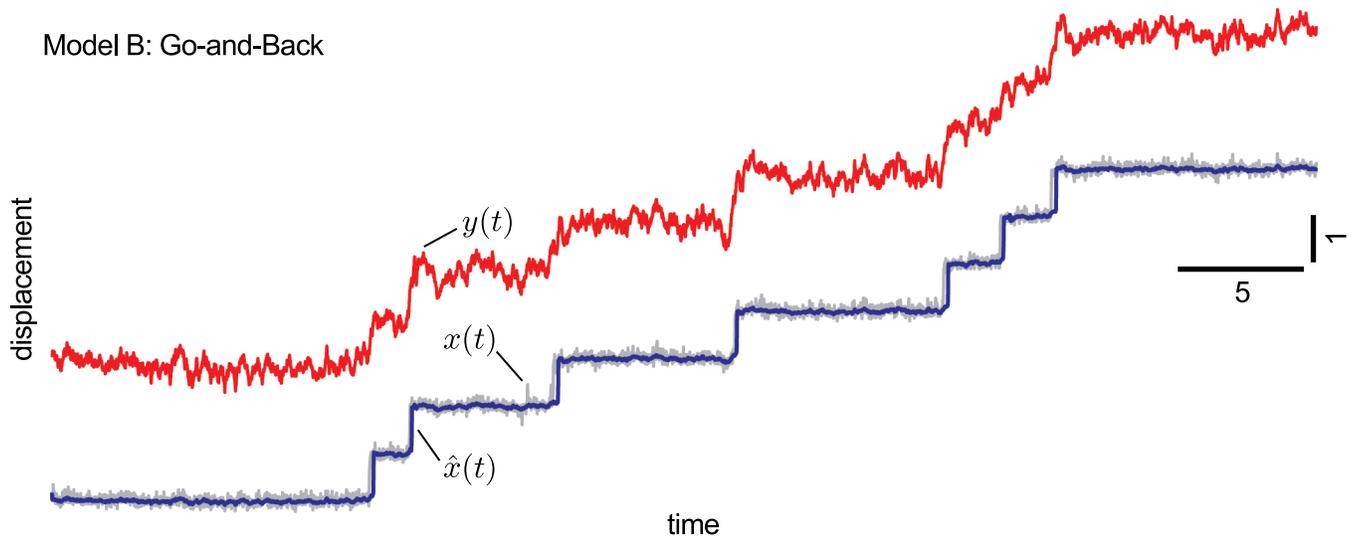}}
\caption{Estimation result of model B by the Go-and-Back method. (top) The trajectory of $y(t)$ [red line]. (bottom) the true trajectory of $x(t)$ [gray line] and the estimated trajectory $\hat{x}(t)$ [blue line]. $i=50$ and $\tau = 50$. For the present parameter setting, the relaxation time scale of $x(t)$ is shorter than $\Delta t$. Therefore, we use linear-interpolated $[y]$ with to stably integrate Eqs.~(\ref{e.gb_x}) and (\ref{e.gb_v}).}
\label{f.tc_step_gb}
\end{figure*}

\section{Examples}
To investigate the practical utility of the Go-and-Back method, we examine the two models illustrated in Fig.~\ref{f.model}. We numerically integrate the model equations [Eqs.~(\ref{e.model_x}) and~(\ref{e.model_y})] from $t =0$ to $t=\tau$ and obtain the true trajectory set $[x,y]$. Then, we assume that we can only monitor $[y]$, and estimate $[x]$ from $[y]$. 

Figures~\ref{f.tc_dw_gb} and~\ref{f.tc_step_gb} display the estimation results of model A and model B by the Go-and-Back method, respectively. Although the estimated trajectory of $x(t)$, denoted as $[\hat{x}]$, does not coincide with the true trajectory $[x]$, $[\hat{x}]$ seems to be a good estimate in both examples. In particular, stepwise motion of $x(t)$ in model B  is precisely reproduced from noisy $y(t)$ (Fig.~\ref{f.tc_step_gb}).

\begin{figure}[tbp]
\centerline{\includegraphics{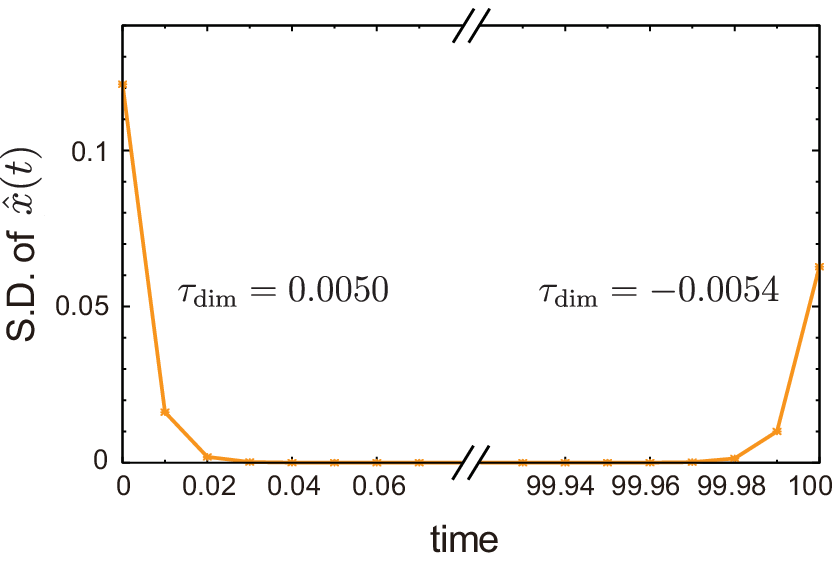}}
\caption{Effect of the boundary condition. The boundary condition $\{x_{(i)}(0), v_{(i)}(\tau)\}$ of model A is varied randomly  ($n=20$), and the standard deviations of $\hat{x}(t)$ at each $t$ are plotted. The diminishing times at both boundaries are evaluated by fitting exponent functions. $x_{(i)}(0)$ is varied Gaussian noise with mean $y(0)$, S.D. $=0.5$ [half length of the two stable states]. $v_{(i)}(\tau)$ is varied zero-mean Gaussian noise with the variance $2\gamma\Tm$ [the same power of $\xi(t)$].}
\label{f.BC}
\end{figure}

To apply the Go-and-Back method, a boundary condition is required. (The initial condition is included in the algorithm [Eq.~(\ref{e.v_0})]. The validity well be discussed below.) For both examples (Figs.~\ref{f.tc_dw_gb} and \ref{f.tc_step_gb}), $\{ x_{(i)}(0), v_{(i)}(\tau)\} = \{y(0), 0\}$ is adopted. To investigate the stability of the Go-and-Back method, we vary the boundary condition of model A at random. As a result, effect of the boundary condition is diminished instantly (Fig.~\ref{f.BC}). Only less than 0.1\% of the total length of the trajectory at both ends is affected by the boundary condition. Thus, the estimates are almost uniform against the choice of the boundary condition. The Go-and-Back method allows to expect the diminishing time scale from Eq.~(\ref{e.led_g}) by the second order approximation on $G(x, y)$ around the bottom of the effective potential. For the present model, the diminishing time scale is approximated as $\tau_{\rm dim} \sim 0.001$,\footnote{$\tau_{\rm dim} \sim \gamma / G_{xx}(x_{\rm b}) = \gamma/(a_1 + k)$ where $x_{\rm b}$ represents the bottom position of the effective potential.} which is almost consistent with the numerical results (Fig.~\ref{f.BC}).

\begin{figure}[tbp]
\centerline{\includegraphics{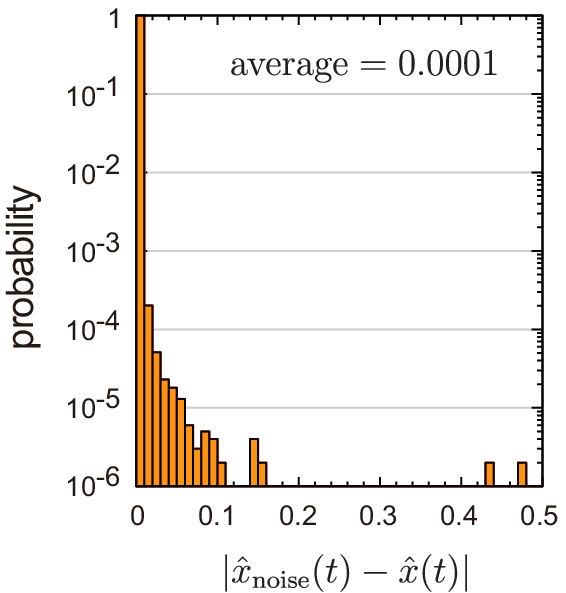}}
\caption{Effect of the initial condition. Zero-mean white Gaussian noise with the variance $0.25\times2\gamma\Tm$ is added to the original initial condition of the Go-and-Back method [Eq.~(\ref{e.v_0})], and the difference between the estimate from the noise-added initial condition $\hat{x}_{\rm noise}(t)$ and the the original estimate $\hat{x}(t)$ is evaluated at each $t$ ($10^{4}$ data points). See also Fig.~S1.}
\label{f.IC}
\end{figure}

We also examine the validity of the initial condition adopted in the algorithm. We apply noise to the original initial condition [Eq.~(\ref{e.v_0})] and evaluate the dependency on the estimate. As a result, the original estimate and the estimate obtained from the noise-added initial condition are almost completely overlapped (Fig.~\ref{f.BC} and Fig.~S1). The difference is $10^{-4}$ on average, which is $10^4$ times smaller than the distance between two stable states, and $10^3$ times smaller than the thermal fluctuation of $x(t)$.\footnote{The standard deviation of the thermal fluctuation of $x(t)$ is $\sqrt{2\Tm \Delta t / \gamma}$ in the case of discrete $t$. For model A, S.D. $= 0.2$.} We note that the Go-and-Back method bases on the Euler-Lagrange equation [Eq.~(\ref{e.E-L})], which means we cannot guarantee that the obtained estimate is the global minimum. However, such a robustness against the choice of the initial condition suggests that the estimate may be the global optimum solution at least in the present model. To conclude, the Go-and-Back method incorporates the appropriate initial condition that yields stable solution, and the method is quite robust against the boundary condition.

\begin{figure}[tbp]
\centerline{\includegraphics{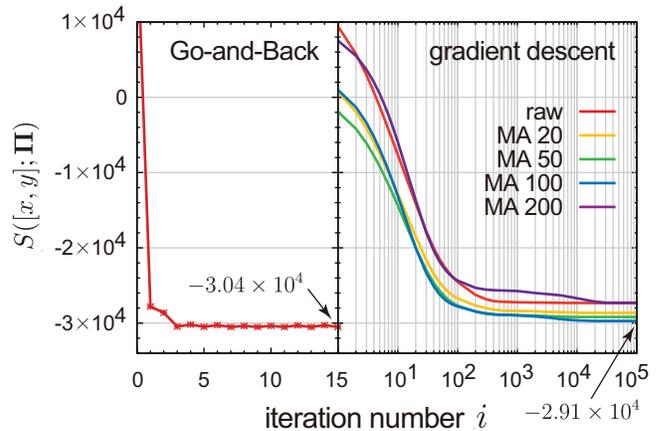}}
\caption{Relaxation properties of the action functional $S([x,y] ; \vec \Pi)$  in the optimization process. (left) Go-and-Back method. The data is taken from the numerical experiment shown in Fig.~\ref{f.tc_dw_gb} (model A). (right) Gradient descent method. We use the same $[y]$ as for the Go-and-Back method (Fig.~\ref{f.tc_dw_gb}, top). For the initial condition, we test raw data of $[y]$ and the moving-averaged (MA) trajectories of $[y]$ with bin sizes 20, 50, 100, and 200. (The time lengths are 0.2, 0.5, 1, and 2, respectively.)}
\label{f.s_dw}
\end{figure}

Next, we compare the Go-and-Back method with the conventional gradient descent method. We use the same $[y]$ as for the Go-and-Back method (Fig.~\ref{f.tc_dw_gb}, top) and adopt the FTCS (forward-time centered-space) scheme\cite{Press:2007_chap20} for the minimization of $S([x,y];\vec \Pi)$. To apply FTCS scheme, the initial and the boundary conditions are required. For the boundary condition, we adopt Dirichlet condition $\{\hat{x}_0, \hat{x}_\tau\} = \{y_0, y_{\tau}\}$, and for the initial condition, we adopt raw data of $[y]$, or the moving-averaged trajectories of $[y]$ with different bin sizes in order to investigate effect of the initial condition.

\begin{figure*}[t]
\centerline{\includegraphics{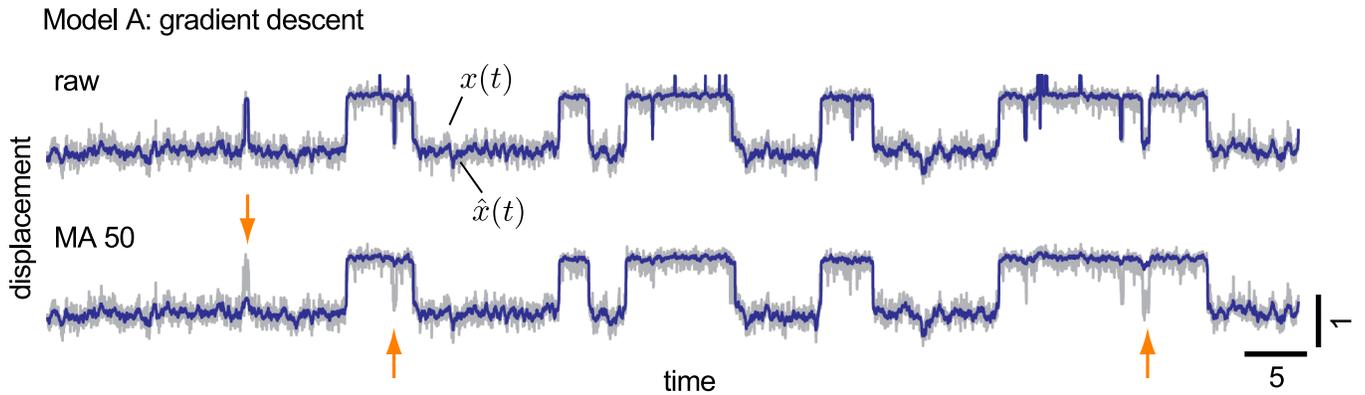}}
\caption{Two examples of the estimation results of model A by the gradient descent method. The true trajectories of $x(t)$ [gray line], and the estimated trajectories $\hat{x}(t)$ [blue line] are shown. (top) The raw data of $[y]$ is adopted for the initial condition. (bottom)
A moving-averaged trajectory of $[y]$ with bin size 50 is adopted for the initial condition. The orange arrows show the regions where $x(t)$ cannot be precisely estimated. In both figures, $i=10^5$.}
\label{f.tc_dw_ftcs}
\end{figure*}

Figure~\ref{f.s_dw} shows the relaxation property of $S([x,y];\vec \Pi)$ between the two methods. In the case of the Go-and-Back method, after the quick relaxation the value of $S$ fluctuates slightly around the minimum value. In some cases, we can recognize that the method overcomes large barrier of $S([x,y] ; \vec \Pi)$ (Figs.~S2 and S3).\footnote{The Go-and-Back method is not an algorithm to search the global minimum point of $S([x,y] ; \vec \Pi)$ along $[x]$-space, but systematically calculates the higher-order approximation of the solution of the Euler-Lagrange equation [Eq.~(\ref{e.E-L})]. Therefore, a monotonic decrease of $S([x,y] ; \vec \Pi)$ does not mean that the estimate gets stuck into the local minimum.} For the present model, $i=15$ is enough to converge the solution. Compared to the Go-and-Back method, as one expects, the gradient descent method requires too many iteration steps for the relaxation. (The relaxation time is $10^4$ times slower.) Moreover, the converged value of $S([x,y]; \vec \Pi)$ in the case of the gradient descent method is strongly dependent on the initial condition. In the present case, roughly smoothed data (MA 50) yields the smallest value of $S([x,y];\vec \Pi)$, but it is still a little bit larger than that of the Go-and-Back method. Such a result implies that the optimization processes are trapped at the local minima. Indeed, $\hat{x}(t)$ is varied among the initial conditions. Two examples are shown in Fig.~\ref{f.tc_dw_ftcs}. Even the best optimized solution of the gradient descent method (MA 50), the motion of $x(t)$ cannot be precisely estimated (Fig.~\ref{f.tc_dw_ftcs}, bottom).
 
We also apply the gradient descent method to model B and obtain features similar to those observed for model A (Figs.~S3 and S4).

\section{Conclusion and Discussion}
We developed a method to estimate the most probable trajectory of the hidden variable from the trajectory of the probe particle. The method naturally incorporates Langevin dynamics. Therefore, although the difficulty of the model settings still remains, if we carefully choose the model, we may extract much more information from experimental data than using the conventional step analysis.

By use of simple models of single-molecule experiments, we numerically verify that the proposed method provides us reasonable estimates. Comparing to the conventional gradient descent method, our proposed method can successfully reduce the computational cost more than $10^3$-fold. In the case of the gradient descent method, the choice of the initial conditions is crucial for obtaining accurate estimates: the wrong choice leaves the optimization process trapped at the local minima. Although we can adopt several techniques, for example, simulated annealing, to overcome this problem, case-by-case treatment is required to improve efficiency. In contrast, our method naturally incorporates an appropriate initial condition. Besides the simplicity of our coding, the proposed method only requires a boundary condition, and the results are quite robust with respect to this choice. Therefore, our method is easy to use, which is important for general users.

As we mentioned above, the present model has several limitations to apply actual experiments. In the present model, it is assumed that the friction coefficient of protein denoted by $\gamma$ is independent of $x(t)$. This term originates from the internal friction of protein and the viscous friction between the protein and the medium. In general, the internal friction should be varied along the reaction coordinate. Indeed, position-dependent coefficients have been obtained from model systems and several kinds of proteins by simulation.\cite{Hummer:2005b, Best:2006, Yang:2006, Best:2010} Therefore, although the constant approximation is not valid for all kinds of proteins, our assumption may be proper for the protein that has a large globular domain and the entire domain tilts or shifts in its structural transition because the internal friction is negligible compared to the large viscous friction. 

In addition, the present model assumes that the effective energy potential of the protein $V^{\rm eff}(x)$ is time independent. However, recent single-molecule studies on cholesterol oxidase\cite{Lu:1998} and $\beta$-galactosidase\cite{English:2005} showed direct evidence that the enzymatic turnover events are not a Markovian process but correlated with the previous history. The result indicates that the protein has slow conformation fluctuations and this time dependence should be appeared in $V^{\rm eff}(x)$. Although, as far as we know, the similar slow dynamics has never been observed in the motor proteins, one has to verify in advance whether such a slow dynamics presents in the protein of interest. By contrast, if the substrate of the enzymatic reaction is not abundant in the medium so that the rate limiting step of the turnover is substrate binding, at least, on and off states should be incorporated into the model. Namely, $V^{\rm eff}(x)$ should have two (or more) states and switch stochastically.\cite{Julicher:1997, Reimann:2002, Harada:2005a, Harada:2005b} For this case, new efficient estimation algorithm must be developed on the basis of the switching state model.

In actual applications of our model to experiments, we must estimate the parameters of the entire model in advance. We have proposed a general framework of parameter estimation in the presence of hidden degrees of freedom.\cite{Miyazaki:2011} According to this framework, which is based on Bayesian statistics, we can precisely estimate the model parameters by maximizing the marginalized path probability with respect to the parameters. The marginalized path probability, called a marginal likelihood, is calculated by integrating all possible trajectories of the hidden variables. In this calculation, the DRP, or the MAP estimator in the language of Bayesian statistics, plays the central role. For the analytical approach, the Wentzel--Kramers--Brillouin (WKB) approximation around the MAP estimator may work well when the temperature of the system is sufficiently small.\cite{Hunt:1981, Miyazaki:2011} For the numerical approach, we can utilize the MAP estimator for the initial condition of the Markov-chain Monte Carlo (MCMC) method\cite{Bishop:2006, Dellago:1998}. By the use of the proposed method, the next step is to investigate how effectively we can obtain reasonable estimates of the model and the hidden trajectories simultaneously.

\begin{acknowledgments}
The authors thank M.~Y.~Matsuo and S.~Toyabe for fruitful discussions. The authors also thank M.~Opper, M.~Sano, M.~Ichikawa, and K.~Yoshikawa for helpful comments, and K.~Shiroguchi, M.~Sugawa, T.~Nishizaka, and T.~Okamoto for discussions on single-molecule experiments. This work was supported from the JSPS Research Fellowships for Young Scientists, No.~20-4954 (to M.~M.), and a grant from MEXT, No.~20740239 (to T.~H.).
\end{acknowledgments}

\appendix

\section{Perturbation expansion}
We expand $x(t)$ and $v(t)$ as
\begin{eqnarray}
	x(t) &=& \sum_{i=0}^{\infty} \varepsilon^{i} x^{(i)}(t), \label{e.x_pe} \\
	v(t) &=& \sum_{i=0}^{\infty} \varepsilon^{i} v^{(i)}(t), \label{e.v_pe}
\end{eqnarray}
where we denote the $i$-th order terms of $x(t)$ and $v(t)$ as $x^{(i)}(t)$ and $v^{(i)}(t)$, respectively, in order to distinguish from $x_{(i)}(t)$ [Eq.~(\ref{e.x_purt})] and $v_{(i)}(t)$ [Eq.~(\ref{e.v_purt})].

The $0$-th order terms become
\begin{eqnarray}
	\gamma \dot{x}^{(0)} &=& -G_x(x^{(0)}, y) + v^{(0)}, \label{e.x_pe_0} \\
	\gamma \dot{v}^{(0)} &=& G_{xx} (x^{(0)}, y) \ v^{(0)}.  \label{e.v_pe_0}
\end{eqnarray}
$G_{xx} (x^{(0)}, y) > 0$ in typical cases, and thus $v^{(0)} (t)$ will instantly destabilize when we solve Eq.~(\ref{e.v_pe_0}) in the forward-time direction. In contrast, if we solve Eq.~(\ref{e.v_pe_0}) in the reverse-time direction, it is expected that $v^{(0)} (t)$ quickly relaxes to around $0$. Therefore, we can naturally assume the approximation $v^{(0)}(t) = 0$.  Since $v^{(0)}(t)$ is identical to $v_{(0)}(t)$, we introduce the approximation Eq.~(\ref{e.v_0}) in the main text. 

Owing to this assumption, we can stably solve $x^{(0)}(t)$ in the forward direction by adopting the initial condition on $x^{(0)}(0)$.

The first-order terms become
\begin{eqnarray}
	\gamma \dot{x}^{(1)} &=& -G_{xx} (x^{(0)}, y) x^{(1)} + v^{(1)}, \label{e.x_pe_1} \\
	\gamma \dot{v}^{(1)} &=& G_{xx} (x^{(0)}, y) v^{(1)} - \Tm G_{xxx} (x^{(0)}, y). \label{e.v_pe_1}
\end{eqnarray}
It is noteworthy that $x^{(1)}(t)$ is not included in Eq.~(\ref{e.v_pe_1}), and thus we can stably solve Eq.~(\ref{e.v_pe_1}) in the reverse direction. Then, by substituting $v^{(1)}(t)$ into Eq.~(\ref{e.x_pe_1}), we obtain $x^{(1)}(t)$.

Similarly, the second-order terms become
\begin{eqnarray}
	\gamma \dot{x}^{(2)} &=& -\frac{1}{2}G_{xxx} (x^{(0)}, y) [x^{(1)}]^2 - G_{xx} (x^{(0)}, y) x^{(2)} \nonumber \\
	&& + v^{(2)}, \label{e.x_pe_2} \\
	\gamma \dot{v}^{(2)} &=& G_{xxx} (x^{(0)}, y)  x^{(1)} v^{(1)}  + G_{xx} (x^{(0)}, y) v^{(2)} \nonumber \\
	&& - \Tm G_{xxxx} (x^{(0)}, y) x^{(1)}. \label{e.v_pe_2}
\end{eqnarray}
Again, $x^{(2)}(t)$ is not included in Eq.~(\ref{e.v_pe_2}). Therefore, we obtain $v^{(2)}(t)$ and $x^{(2)}(t)$ by the use of the lower-order solutions.
In this way, we can systematically calculate the higher-order terms.

However, as the order number increases, the analytical solution becomes complicated. We further develop the method to simplify the algorithm.

We introduce Eqs.~(\ref{e.x_purt}) and~(\ref{e.v_purt}). In these definitions, $x_{(0)}(t)$ and $v_{(0)}(t)$ are identical to $x^{(0)}(t)$ and $v^{(0)}(t)$, respectively. 

Next, combining Eqs.~(\ref{e.x_pe_0})  and~(\ref{e.x_pe_1}) leads to
\begin{eqnarray}
	\gamma \dot{x}_{(1)} &=& -G_x(x_{(1)}, y) + v_{(1)} + O(\varepsilon^2), \label{e.x_purt_1}
\end{eqnarray}
and combining Eqs.~(\ref{e.v_pe_0}) and~(\ref{e.v_pe_1}) leads to
\begin{eqnarray}
	\gamma \dot{v}_{(1)} &=& G_{xx} (x_{(0)}, y) v_{(1)} - \varepsilon \Tm G_{xxx} (x_{(0)}, y). \label{e.v_purt_1}
\end{eqnarray}
Similarly,
\begin{eqnarray}
	\gamma \dot{x}_{(2)} &=& -G_x(x_{(2)}, y) + v_{(2)} + O(\varepsilon^3), \label{e.v_purt_2} \\
	\gamma \dot{v}_{(2)} &=& G_{xx} (x_{(1)}, y) v_{(2)} - \varepsilon \Tm G_{xxx} (x_{(1)}, y) \nonumber \\
	&& + O(\varepsilon^3). \label{e.x_purt_2}
\end{eqnarray}
In this manner, we obtain the general forms, Eqs.~(\ref{e.gb_x}) and~(\ref{e.gb_v}).



\providecommand{\noopsort}[1]{}\providecommand{\singleletter}[1]{#1}%

\end{document}